\documentclass[a4paper,11pt]{article}
\usepackage{pos}
\usepackage{multirow}

\title{Feasibility study of an accelerator neutrino experiment in China}
\author{Jian Tang}
\author*{Sampsa Vihonen}
\author{Yu Xu}

\affiliation{School of Physics, Sun Yat-sen University,\\
  No. 135 Xingang Xi Road, 510275 Guangzhou, P.R. China}

\emailAdd{sampsa@mail.sysu.edu.cn}

\abstract{Future accelerator neutrino experiments will provide a powerful tool to measure standard oscillation parameters and search for new physics. In this context, we discuss the prospects of building an accelerator neutrino experiment in China. The feasibility of such facilities is investigated by evaluating their prospects to the standard mixing parameters. As an example, we consider an SPPC-based neutrino beamline and CJPL-based neutrino detector with 1736~km baseline length. We find this setup able to significantly improve the precision on $\delta_{CP}$, $\theta_{23}$ and $\Delta m_{31}^2$.}

\FullConference{%
  *** The European Physical Society Conference on High Energy Physics (EPS-HEP2021), ***\\
  *** 26-30 July 2021 ***\\
  *** Online conference, jointly organized by Universit\"at Hamburg and the research center DESY ***
}

\begin{document}
\maketitle

\section{Introduction}
\label{sec:intro}

China has rapidly growing infrastructure dedicated to accelerator-based nuclear and particle physics~\cite{Tang:2021lyn}. China Spallation Neutron Source (CSNS), China initiative for Advanced Driven Systems (CiADS) and High Intensity Accelerator Facility (HIAF) in the southern province of Guangdong, for example, are major accelerator facilities with applications ranging from neutron scattering techniques to nuclear transmutation and heavy ion physics. Institute of Modern Physics of Chinese Academy of Sciences (CAS-IMP) in Lanzhou, Gansu, and Proton Linear Accelerator Institute of Nanjing University in Nanjing, Jiangsu, are other notable Chinese laboratories specialized in accelerator-based physics. At the same time, a next-generation collider facility CEPC and its high-energy upgrade SPPC are planned to be constructed in Beijing in the northeast of China. Last but not least, China is home to two major underground physics laboratories designed for investigating fundamental sciences: China JinPing Laboratory (CJPL) in central Sichuan is currently the deepest underground facility in the world and Jiangmen Underground Neutrino (JUNO) Observatory will soon commission the largest liquid scintillator neutrino detector near Kaiping, Guangdong.

In this proceeding, we briefly contemplate the prospects of studying precision neutrino oscillation physics with a future accelerator neutrino experiment in China. We discuss the site selection aspect with respect to the geographical locations of the existing research infrastructure in China. As a concrete example, we consider an experiment configuration where a neutrino beamline is established at the SPPC injector facility and detector at CJPL, giving rise to 1736~km baseline.

This proceeding is organized as follows: We summarize the key points of the laboratory survey in section\,\ref{sec:labs}, discuss the physics prospects of the considered setups in the precision measurement of $\theta_{23}$, $\delta_{CP}$ and $\Delta m_{31}^2$ in section\,\ref{sec:res} and leave concluding remarks in section\,\ref{sec:concl}. 

\section{Survey of research infrastructure in China}
\label{sec:labs}

China is currently home to a number of accelerator facilities with a range of interests in applied and basic sciences. There are also two major underground research laboratories. If these accelerator and underground laboratories were used in a future accelerator neutrino experiment, the available baseline lengths would range from 162~km to 1871~km. A summary of baseline lengths and required neutrino beam energies to reach the first and second oscillation maximum in these facilities is provided in table~\ref{tab:labsites2}. 
\begin{table}[!ht]
\caption{\label{tab:labsites2}Characteristics of the five accelerator laboratories and two underground laboratories in China. Listed items include the baseline lengths and the required neutrino energy to reach first and second oscillation maximum, assuming $\Delta m_{31}^2 \simeq$ 2.517$\times$10$^{-3}$~eV$^2$. This table is taken from Ref.~\cite{Tang:2021lyn}.}
\begin{center}
\resizebox{0.85\linewidth}{!}{
\begin{tabular}{ |c|c|c|c|c|c|c| } 
 \hline
 \multirow{2}{*}{Accelerator facility} & \multicolumn{3}{ c| }{JUNO} & \multicolumn{3}{ c| }{CJPL} \\
 \cline{2-7}
 & Baseline & 1$^{\rm st}$ maximum & 2$^{\rm nd}$ maximum & Baseline & 1$^{\rm st}$ maximum & 2$^{\rm nd}$ maximum \\
 \hline
 CAS-IMP & 1759~km & 3.6~GeV & 1.2~GeV & 894~km & 1.8~GeV & 600~MeV \\ \hline
 CiADS & 221~km & 450~MeV & 150~MeV & 1389~km & 2.8~GeV & 940~MeV \\ \hline
 CSNS & 162~km & 330~MeV & 110~MeV & 1329~km & 2.7~GeV & 900~MeV \\ \hline
 Nanjing & 1261~km & 2.6~GeV & 850~MeV & 1693~km & 3.4~GeV & 1.1~GeV \\ \hline
 SPPC & 1871~km & 3.8~GeV & 1.3~GeV & 1736~km & 3.5~GeV & 1.2~GeV \\
 \hline
\end{tabular}}
\end{center}
\end{table}

\begin{figure}[!t]
        \center{\includegraphics[width=0.9\textwidth]{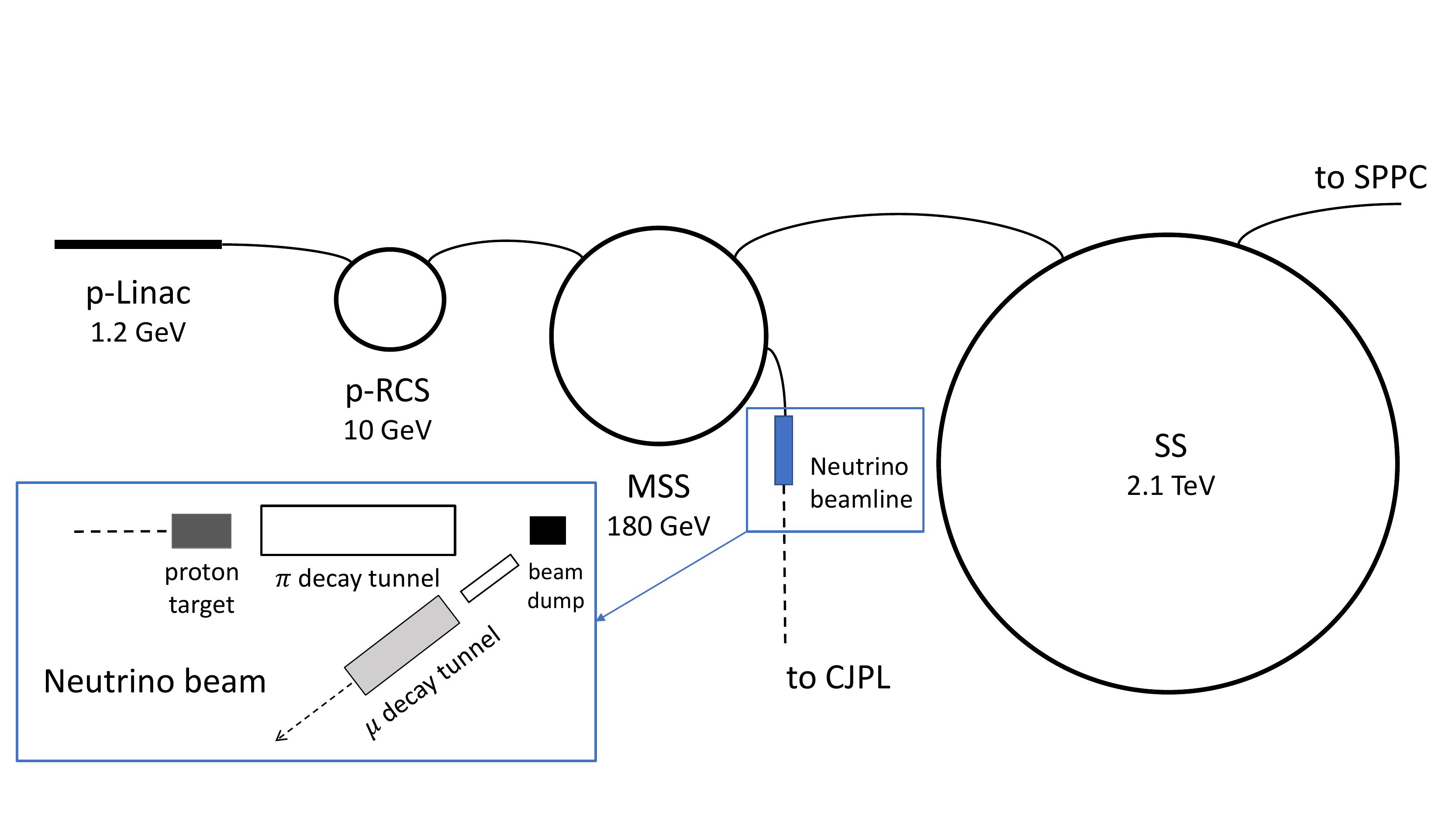}}
        \caption{Illustration of a possible realization of neutrino beamline at the SPPC injector facility~\cite{Tang:2021lyn}. Both MSS and SS rings could divert an average of 3.2~MW proton beam to non-collider purposes~\cite{CEPC-SPPCStudyGroup:2015esa}.}
        \label{fig:configuration}
\end{figure}

An attractive candidate for a future neutrino beam facility site is the SPPC injector chain, as portrayed in figure~\ref{fig:configuration}. SPPC is part of the next-generation collider physics program CEPC-SPPC, which is currently under planning in China~\cite{CEPC-SPPCStudyGroup:2015esa}. The injector chain consists of a proton linac (p-Linac), rapid  cycling  synchrotron (p-RCS), medium-stage  synchrotron (MSS) and final  stage  synchrotron (SS). Each synchrotron is capable of diverting a proton beam of 3.2~MW to non-collider programs.

\section{Precision measurements on $\delta_{CP}$, $\theta_{23}$ and $\Delta m_{31}^2$ with PROMPT}
\label{sec:res}

In this proceeding, we present the expected sensitivities for an accelerator neutrino experiment based at the SPPC injector facility in Beijing and CJPL in Sichuan. For future reference, we call this setup {\em PRecisiOn Measurements and Physics with Tau neutrinos} (PROMPT) due to its affinity to $\nu_\tau$ physics and perceived physics goals. PROMPT is simulated as a long-baseline neutrino experiment with muon-decay-based beam of 25~GeV parent energy and 50~kton hybrid detector based on magnetized iron and emulsion cloud chamber technologies. 

\begin{figure}[!t]
        \center{\includegraphics[width=0.65\textwidth]{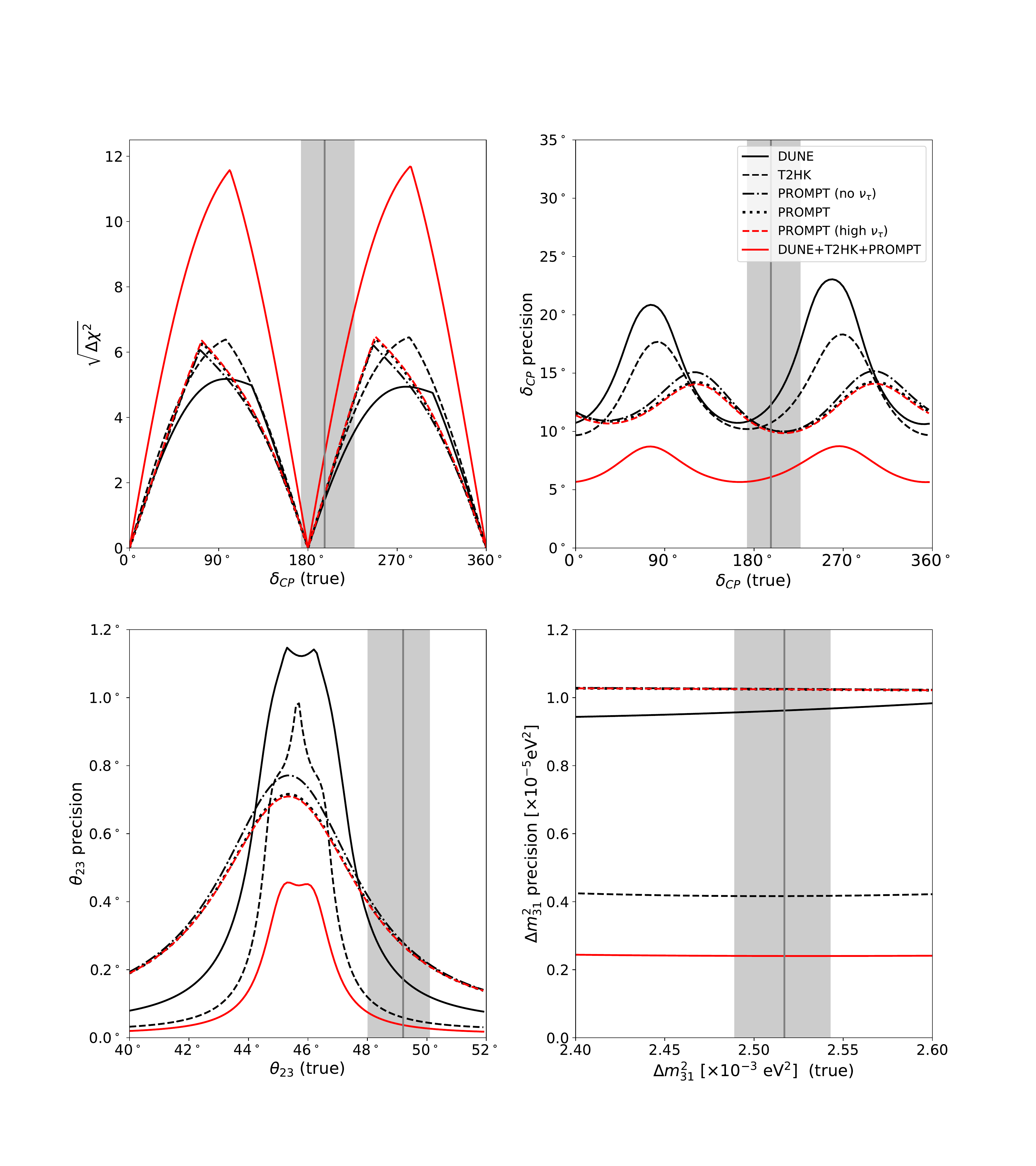}}
        \caption{Expected sensitivities to {\em CP} violation and 1$\,\sigma$ CL precision to $\theta_{23}$, $\delta_{CP}$ and $\Delta m_{31}^2$ at PROMPT setup~\cite{Tang:2021lyn}. Precisions are shown with respect to the presently favoured value from a global analysis, which are also indicated within 1$\,\sigma$ CL by gray colour. Sensitivities from T2HK and DUNE are also provided.}
        \label{fig:precision}
\end{figure}

The expected sensitivities for {\em CP} violation discovery as well as for precision measurement of $\theta_{23}$, $\delta_{CP}$ and $\Delta m_{31}^2$ are shown in figure~\ref{fig:precision}. The simulations are conducted with GLoBES. The sensitivity to {CP} violation is presented as a function of $\delta_{CP}$ values, whereas the precisions on $\theta_{23}$, $\delta_{CP}$ and $\Delta m_{31}^2$ are indicated with respect to the values presently favoured by the world data~\cite{Esteban:2020cvm}. The best-fit values are also shown at 1$\,\sigma$ CL with grey regions. The sensitivities for PROMPT are shown three different scenarios for $\nu_\tau$ efficiency, see Ref.~\cite{Tang:2021lyn} for more details. The corresponding sensitivities expected for DUNE and T2HK as well as the combination of the three configurations are also included. The sensitivities predicted for PROMPT show excellent precision to $\delta_{CP}$, $\theta_{23}$ and $\Delta m_{31}^2$ both as a standalone experiment and in combination of the simulated data from the superbeam experiments T2HK and DUNE.

\section{Summary and outlook}
\label{sec:concl}

China offers a great landscape in research infrastructure dedicated to nuclear and particle physics. In this proceeding, we contemplated the possibility of constructing a future accelerator neutrino experiment in China. As a concrete example, we considered an experiment setup at the SPPC and CJPL sites, which could form a baseline length of 1736~km. We present the expected sensitivities to {\em CP} violation and the precision of $\theta_{23}$, $\delta_{CP}$ and $\Delta m_{31}^2$ in this setup, which we call PROMPT. We find that PROMPT could significantly improve the experimental precision on the standard mixing parameters.

%\acknowledgments
%This work was supported supported in part by National Natural Science Foundation of China under Grant No. 12075326 and No. 11881240247, Guangdong Basic and Applied Basic Research Foundation under Grant No. 2019A1515012216 and China Postdoctoral Science Foundation under Grant No.\,2020M672930.


\begin{thebibliography}{99}
\bibitem{Tang:2021lyn}Jian Tang, Sampsa Vihonen and Yu Xu, \href{https://arxiv.org/abs/2108.11107}{arXiv:2108.11107}.
\bibitem{CEPC-SPPCStudyGroup:2015esa}CEPC-SPPC Study Group, IHEP-CEPC-DR-2015-01, IHEP-AC-2015-01 (2015).
\bibitem{Esteban:2020cvm}Ivan Esteban, M.C. Gonzalez-Garcia, Michele Maltoni, Thomas Schwetz and Albert Zhou, JHEP {\bf 09} (2020) 178, \href{https://arxiv.org/abs/2007.14792}{arXiv:2007.14792}.
\end{thebibliography}
\end{document}